\documentclass[10pt,preprint2]{aastex}

\newcommand{\eb}{\begin{equation}}
\newcommand{\ee}{\end{equation}}

\newcounter{arabnum}
\newcommand{\befigcap}{\begin{list}{ {\bf Figure \arabic{arabnum} } %
{ \usecounter{arabnum}} } }
\newcommand{\enfigcap}{\end{list}}

\newcommand{\bequo}{\begin{quotation}}
\newcommand{\enquo}{\end{quotation}}
\newcommand{\bverse}{\begin{verse}}
\newcommand{\everse}{\end{verse}}
\newcommand{\beit}{\begin{itemize}}
\newcommand{\enit}{\end{itemize}}
\newcommand{\been}{\begin{enumerate}}
\newcommand{\enen}{\end{enumerate}}
\newcommand{\ecen}{\end{center}}
\newcommand{\bcen}{\begin{center}}
\newcommand{\begeq}{\begin{equation}}
\newcommand{\eneq}{\end{equation}}
\newcommand{\befig}{\begin{figure}}
\newcommand{\enfig}{\end{figure}}

\newcommand{\loglx}{\mbox{$\log L_{\it x}$}}
\newcommand{\logl}{\mbox{$\log L$}}

\newcommand{\uvby}{\mbox{\it uvby }}
\newcommand{\dmczero}{\mbox{$\Delta M_{c_0}$ }}
\newcommand{\tef}{\mbox{$T_e\:$}}
\newcommand{\logtef}{\mbox{$\log T_e\:$}}
\newcommand{\ferrh}{\mbox{$\rm{[Fe/H]\:}$}}

\newcommand{\kmsec}{\mbox{${\rm \: km\:s^{-1}}$}}

\newcommand{\msol}{\mbox{$\: M_\odot$}}

\newcommand{\hipparcos}{\mbox{\it Hipparcos }}
\newcommand{\rosat}{\mbox{\it Rosat}}
\begin{document}

\slugcomment{}

\title {F stars: A challenge to stellar evolution} 

\author{A. A. Suchkov and S. A. Lyapustina
\affil{\medskip Rockland Hill Research Group, 8 Deaven Ct, Baltimore, MD 21209} \email{suchkov.md@gmail.com.} 
}

\begin{abstract}

Many main-sequence F and early G stars are   too luminous for their effective 
temperature, surface gravity, and chemical composition. These
{\it overluminous stars} have two curious properties. First, their kinematics 
as a function of age from
stellar evolution modeling (isochrone fitting) is very different from
that of normal stars.  
Second, while  X-ray luminosity of normal stars declines with age,  
the X-ray luminosity of overluminous F stars changes in the opposite direction, 
being on average higher for older stars. These properties imply that,  
in defiance of standard models of stellar evolution, F stars of a given mass 
and chemical 
composition can evolve very differently. Assuming that the models correctly
describe normal stars, for overluminous F stars they 
predict too young age and the X-ray emission evolving in the direction 
opposite to the actually observed trend. This discrepancy between modeling 
results and observational data suggests that standard stellar evolution 
models and models of stellar activity 
are missing some  important factors, which  makes stellar age 
and predictions for stellar activity from these models problematic. 
The data and literature analysis presented in this paper point to 
a nonuniform rotation of the stellar interior as a plausible key factor 
able to reconcile the divergent trends in age-velocity relationships of 
normal and overluminous F stars and  explain in 
a coherent and self-consistent way the overluminosity phenomenon. 

\end {abstract}


\section{Introduction}

Solar-type stars in the mass range of $\sim$ 1.0 to 1.5~\msol, which are
mostly of the spectral type F, are  uniquely suited for examining issues
of stellar evolution, evolution of stellar activity, and the history of star 
formation in the solar neighborhood. Their particular suitability is
due to the large spread in their evolutionary status and the availability 
of large amounts of varied, pertinent data, as summarized below: 
\begin{itemize}
\item 
F stars are the most numerous type in the {\it Hipparcos} catalog, greatly 
outnumbering all other types of stars with  measured 
trigonometric parallaxes.
\item
For well over ten thousand F~stars, {\it uvby} photometric parameters are
available, which yield star's surface gravity along with temperature, 
luminosity, and metal abundance.
\item
Thousands of these stars have measured radial velocities, which in 
combination with proper motion and parallax from {\it Hipparcos} yield 
spatial velocities.
\item
Over two thousand of the same population have {\it Rosat} data, which 
translate to X-ray luminosities when combined with {\it Hipparcos} parallaxes.
\item
Single F stars are well differentiated from binary stars.
\item
F stars on and above the zero-age main sequence (ZAMS) straddle the entire age
 range of stars in the Galactic disk. With known distances and photometry, 
this means the availability of their age---from standard stellar evolution
modeling---in the range of 0 to $\sim\!10$~Gyr. 
\end{itemize}

The abundance of data from {\hipparcos} (ESA 1997), {\rosat} (Voges et al. 
1999, 2000) and  \uvby photometry (Hauck \& Mermillod  1998), with
derived parameters such as temperature and metallicity and computed parameters
such as age, provides a tremendously rich background to explore the evolution 
of stars and stellar  activity of solar-type stars over a period stretching 
from the present epoch back to the time when the first stars 
were formed in the nearby Galactic disk. Earlier, we used the indicated data 
to address some aspects of these issues  
(Suchkov \& McMaster 1999, Suchkov 2000, 2001, Griffin \& Suchkov 2003, 
and Suchkov,  Makarov, \& Voges 2003). More recently Holmberg, Nordstrom,
\& Andersen (2009) published
refined \hipparcos parallaxes along with new data and derived parameters for
many \hipparcos stars. These newly measured parameters were radial velocity and
 multiplicity (differentiation between single, binary, etc. stars). The derived
parameters included temperature, metallicity, spatial (3D) velocity,
and age. In this paper, we take advantage of that body of data to revisit
and expand on our previous work.

\section{Data} 

The base sample used in this paper is built from a set of 11900 
F~stars described in Suchkov \& McMaster (1999). The astrometry and stellar 
parameters are taken from Holmberg et al. (2009).  
Along with the parameter constraints used in Suchkov \& McMaster (1999), 
the current sample is additionally constrained specifically with respect 
to the parameters from Holmberg et al. (2009): 
\begin{itemize}
\item
Parallax error: $\sigma_{\pi}/\pi < 0.15$. 
\item
Distance: $d < 150$~pc.
\item
Spatial velocity: $v < 200$~\kmsec.
\item
Metallicity: $-1.0 < \ferrh < 0.5$.
\item
Multiplicity: no binary or triple stars; also 
no open cluster members.
\end{itemize}
The overluminosity parameter discussed below is constrained by
$-0.15 \leq \dmczero \leq 1.0$.
The subsample  of X-ray emitters has X-ray luminosity from Suchkov, 
Makarov, \& Voges (2003) and is limited to $26 < \loglx  < 33$. 

\section{Overluminosity} 

Suchkov \& McMaster (1999) introduced an {\it overluminosity parameter} 
defined as the difference between absolute magnitude $M_{c_0}$ derived 
from the \uvby photometry and absolute magnitude $M_V$ based on 
\hipparcos parallax:
\begeq
\dmczero = M_{c_0} - M_V.
\eneq
For F stars, the \uvby parameter $c_0$ is, in fact, sensitive to surface
gravity rather than luminosity.
It is a usable measure of luminosity as long as there is a one-to-one
relationship
between surface gravity and luminosity. The reasonably tight relationship
between $c_0$ and $M_V$ found for well-studied nearby single F~stars 
implies that for many stars this is indeed so. But in general, such
a relationship cannot be presumed, and $\Delta M_{c_0}$ differentiates 
single  F~stars according to their luminosity at a given surface gravity, 
effective temperature, and chemical composition.  

By definition, single stars of normal 
luminosity will have   $\Delta M_{c_0} =0$. Given the uncertainty in the 
measured values of the overluminosity parameter, the actual
 operational criterion of normal stars was set to $\Delta M_{c_0} =0 \pm 0.15$,
 which will also be adopted in this paper. 
{\it Overluminous stars} (the term introduced
by Griffin \& Suchkov 2003), i.e., stars whose luminosity 
is higher than that of normal stars of the same  temperature, gravity, 
and chemical composition, are then defined as  those with 
$\Delta M_{c_0} > 0.15$. At a given luminosity and temperature an overluminous
star has a larger surface gravity; hence, of the two stars located at the 
same point in the $\logtef$--$\logl$ diagram an overluminous star  
is more massive than a normal one.

\begin{figure}[ht]
\epsscale{1.0}
\plotone{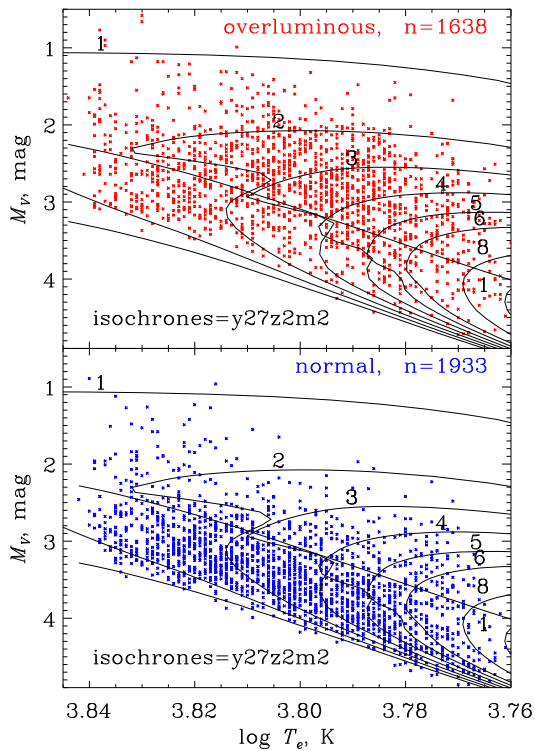}
\caption{$\log T_e - M_V$ diagram for normal and overluminous F stars.
The number of stars is shown in the upper right corner.
A set of isochrones (labeled by age in Gyr), based on Chaboyer at al. 1999 
for composition $Y=0.27$ and $Z=2\times10^{-2}$, and a line one magnitude 
above the ZAMS are shown for reference. 
Overluminous stars tend to avoid the area near the ZAMS and 
are on average brighter than normal stars at the same temperature.
}
\label{f1} 
\end{figure}

Overluminosity distribution is continuous, peaking
at $\dmczero \approx 0$ and extending up to $\dmczero \sim 1.5$ 
(see Figure~1 in Suchkov 2001). The splitting of
F~stars into two groups, normal and overluminous, is just a convenient
way to present the issues associated with the overluminosity phenomenon rather
than a division into two distinct, separate populations.
As argued below, overluminosity reflects peculiarities in stellar evolution
of  stars intrinsically different from normally evolving stars.
In the course  of stellar evolution, the \dmczero of stars
that become overluminous progressively increases at a rate that depends 
on the original difference between these stars and normally evolving stars,
so at any given value of \dmczero there is a mixture of stars of all ages. 
The way we split our sample into two groups ensures that a substantial
fraction of the overluminous group consists of far evolved stars.
This explains why the overluminous group looks so different from 
the normal group in the $\log T_e - M_V$ diagram (Figure~\ref{f1}).

Originally, overluminous stars that were listed as single in the \hipparcos
catalog were believed to be mostly unidentified binary
stars with comparably bright normal components (Suchkov \& McMaster 1999).
An unresolved binary with equally bright normal components 
would look overluminous by $\dmczero = 0.75$. 
The lingering problem  was that there were  
stars with \dmczero significantly larger than 0.75, which definitely did 
not fit to the binary star hypothesis. Another puzzle was that
overluminous stars as a group turned out to be  much older than the 
known binaries in the \hipparcos (Suchkov 2000). So the binaries were not 
the complete answer, and something else must be at work. 
Indeed, later we found that overluminous stars were comprised of at least 
three different subgroups (Griffin \& Suchkov 2003). Of those three one 
subgroup were binaries not identified in the \hipparcos and the other subgroup 
were young pre-main-sequence stars. The third group was comprised of truly 
single main-sequence stars. The stars in this last category are the main focus
 of this paper. 

\section{Overluminosity: Age--velocity relation }

\begin{figure}[ht]
\epsscale{1.0}
\plotone{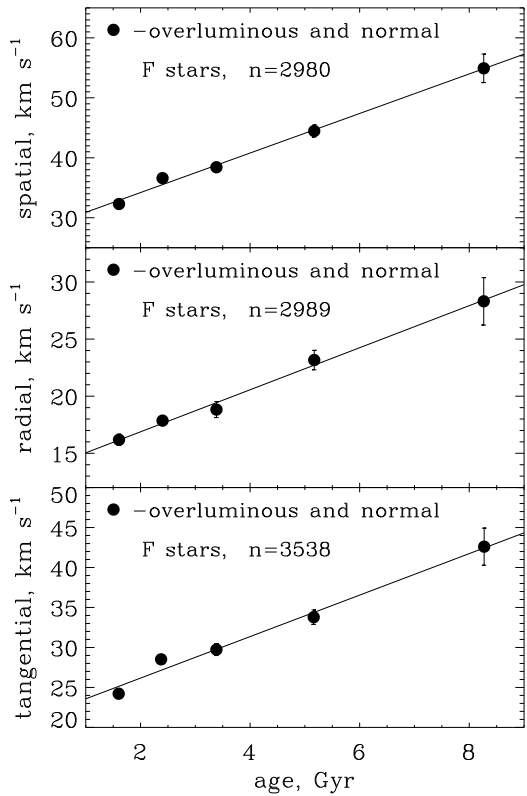}
\caption{Age--velocity relation for F~stars using radial,  
tangential or spatial velocity. The velocity statistics 
is the mean of absolute velocity value. The sample is
limited to stars with $\Delta M_{c_0} > -0.15$.
}
\label{f2} 
\end{figure}

Older stars are long known to have on average higher velocities, which is 
 demonstrated by the age--velocity relation (AVR) of F~stars 
shown in Figure~\ref{f2}.  The effect is usually 
attributed to some kind of a dynamic process that gradually increases 
stellar velocities as time goes on. Thanks to this clear-cut relationship 
between age and stellar velocities, the kinematics parameters 
such as velocity dispersion and the mean of absolute velocities are
commonly used as age markers of stellar populations. If calibrated against 
actual (absolute) physical age (for instance, age based on stellar evolution
models), the resulting ''kinematics'' age
can be used as a substitute for the actual age of physical groups of stars, 
which proves especially useful when direct age determination (more accurately,
age from stellar evolution modeling) is difficult or impossible.

\begin{figure}[ht]
\epsscale{1.0}
\plotone{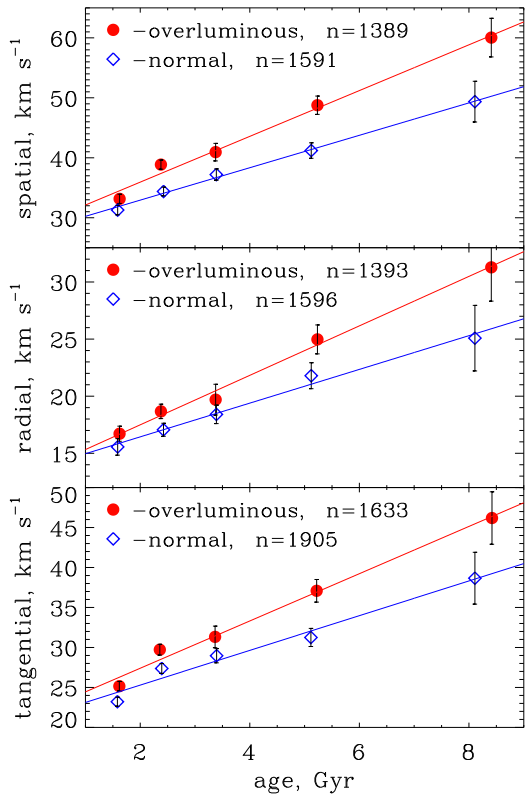}
\caption{Whatever the data source of velocity---Doppler shift (radial velocity),
  parallax plus proper motion (tangential velocity), or the combination of both
 (spatial velocity)---there is a remarkable discrepancy in age-velocity 
relations of normal (circles)  and overluminous (diamonds) F stars, with 
age derived from isochrone fitting. 
}
\label{f3} 
\end{figure}
	
The age parameter in Figure~\ref{f2} is taken from
Holmberg, Nordstrom, \& Andersen 2009. It is based on  ``standard'' stellar 
evolution modeling  that requires initial conditions for only   
three parameters: stellar mass, $M$, helium content, $Y$, and heavy element 
abundance, $Z$.  Standard models provide 
star's luminosity and temperature as  functions of age, thus allowing 
one to compute age from observationally derived luminosity, temperature,
and chemical composition. 
This ''{\it standard}`` age is commonly used in isochrone fitting technique 
to determine  age of   star clusters or individual stars from effective 
temperature, \tef, luminosity, $L$, and chemical composition parameters $Z$ 
and $Y$, so we will also be referring to it as {\it isochrone} age. 

 Depending on how adequate the underlying model is and how  accurate the 
derived stellar parameters are, standard age may differ from the actual age of 
a star. Therefore, if a body of observational data reveals a discrepancy 
between the standard age and actual age, that would  mean that basic premises 
of the standard stellar evolution models need to be reconsidered.

\section{Age and kinematics}

With kinematics depending only on actual (absolute) age, the AVR 
of any group of stars must  be a unique function of {\it standard} 
age if the latter adequately represents the actual age.     
But what if a derived AVR turns out to be a non-unique function?
What if groups of stars differentiated according to some property have 
very different AVR?
The most obvious answer would be that stellar evolution modeling uses 
invalid assumptions, missing  some underlying physics  and/or  some 
parameter(s) that are as important as stellar mass and chemical composition.
Stars differing in such unaccounted for parameter(s)
would produce  different isochrone-based AVRs as their standard age does 
not represent  actual age.

A careful examination of available data demonstrates that F stars do indeed 
reveal a property suggesting that there is some physical factor  missing
from the current standard evolution models, which
 affects stellar evolution in a major way.
The problem can be seen in Figure~\ref{f3} that shows 
age--velocity relation separately for 
normal  and overluminous stars. Instead of being identical, and duplicating
the AVRs in Figure~\ref{f2}, the AVRs for these two subsets
turn out to be quite different. The overluminous stars have
much higher velocities at the same isochrone age, the discrepancy
increasing with age. This means that the actual age of overluminous 
stars is significantly underestimated with respect to normal stars. 
For instance, the kinematics in Figure~\ref{f3} suggests that, at 
isochrone age of 5~Gyr, overluminous stars are in fact $\sim 2.5$~Gyr
 older than normal stars, i.e., older by about 50\%. This inference agrees 
 with our results obtained earlier from our original data set 
 (Suchkov \& McMaster 1999). It is to be noted that the lower panel
 in Figure~\ref{f3} is virtually identical to Figure~7 in Suchkov (2001)
 even though, unlike  Figure~\ref{f3}, the data in the latter Figure 
 are based on the original \hipparcos parallaxes.  

\section{Overluminosity: X-ray -- velocity relation} 

Stellar X-ray emission of late-type stars is a manifestation of stellar 
coronal activity that is associated with the star's outer convection zone. 
The convection zone is thinner 
in stars of larger mass, disappearing in stars earlier than 
spectral type F. This accounts for a stronger coronal activity 
of low-mass stars in comparison with more massive stars of the same age. 
Outer convection apparently disappears altogether in stars earlier than 
$\sim$F5, which is argued to be the reason for the drop of coronal 
activity in this spectral range (e.g., Stauffer et al. 1994).

Stellar X-ray emission is commonly believed to decline with age as coronal
activity decays.
The decrease of X-ray luminosity of main-sequence stars with age was 
inferred back in the 1980's from the {\it Einstein} surveys of open clusters 
and field stars (e.g., Micela et al. 1988). For main-sequence stars, coronal
activity is associated with  magnetic fields generated in the star's  
outer convection zone. Presumably, the field generation mechanism 
(magnetic dynamo) is supported by the combined effect of convection and  
rotation. As the  star ages, rotation slows down and the dynamo 
gets less efficient, which reduces the star's coronal activity and
drives its X-ray emission down.

\begin{figure}[ht]
\epsscale{1.0}
\plotone{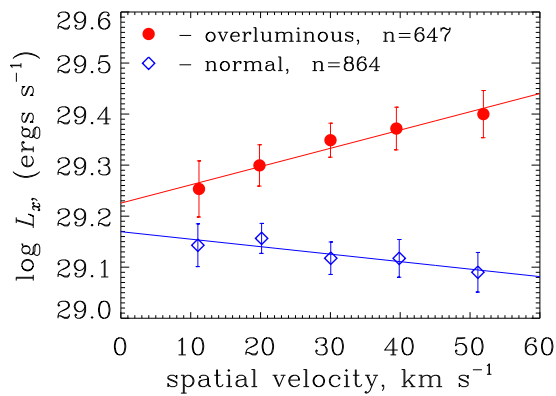} 
\caption{X-ray luminosity of normal F stars (diamonds) slightly 
 declines with increasing velocities, hence with age. In sharp contrast, 
 for overluminous stars (circles), X-ray luminosity trends up 
rather than down, contrary to the conventional picture of 
the stellar activity evolution.    
}
\label{f-f4} 
\end{figure}

Figure~\ref{f-f4} shows the relationship between spatial velocity and 
X-ray luminosity of normal and overluminous F stars.  
Intriguingly, while  the behavior of normal stars in this diagram is 
consistent with the concept of decaying coronal activity, the behavior of
the overluminous stars does not. Overluminous stars are consistently and 
significantly  more luminous at higher velocities than at low velocities, 
which implies  that coronal activity of these stars increases rather than 
decreases with age.  The same conclusion was reached in our earlier study 
(Suchkov, Makarov, \& Voges 2003).

\section{Discussion}

What makes a solar-mass star live longer than the current standard models of 
stellar evolution predict? What can make it get
more active as it ages? What do the standard stellar evolution and stellar
activity modeling miss?
It appears that the only important initial characteristic not included 
in standard modeling, a characteristic that differentiates stars of 
the same mass and chemical composition at the start of their evolution, 
is differential  rotation of 
the stellar interior. Differential rotation of the inner of a star 
can cause radial mixing of the stellar material, transporting fresh
hydrogen fuel into the hydrogen-burning core and thus extending the
star's main-sequence lifetime. Stars with initially different rotational
velocities and velocity radial profiles would experience different amount
of hydrogen enrichment, so their main-sequence lifetimes would be different.

Over the last decade, the probing of internal rotation of stars has
become a major activity in the rapidly developing field of asteroseismology.
As a result, a significant radial variation of rotation in the stellar
interior, from the star's fast rotating core out to the much slower 
rotating upper layers,
 appears now to be a rather well established feature that is typical 
for many stars. 
The respective studies are driven to a great degree by the efforts to 
solve the remaining problems in stellar evolution. 
For instance, Pamyatnykh, Handler, and Dziembowski (2004) addressed 
the problem of the angular momentum evolution and element mixing in 
convectively stable layers.  Using asteroseismology data for 
$\nu$~Eri, they inferred that rotation in the star's
 $\mu$-gradient zone is three times faster than in the envelope.  
 With this kind of velocity gradient one may expect a massive amount 
 of rotationally induced element mixing in the inner layers
 of this star.
 
Stellar evolution modeling that
includes internal rotation as a new factor additional 
to mass and chemical composition is becoming a popular field in stellar 
evolution theory. Eggenberger, 
Maeder, and Meynet (2010) studied the effects of rotational
mixing on the properties of solar-type stars. They concluded that
"Rotational mixing has indeed a large impact on the properties of 
the central layers by bringing fresh hydrogen fuel to the central 
stellar core" and  "Due to rotational mixing, 
the main-sequence lifetime is then larger for stellar
models including rotation" (see also Eggenberger et al. 2010).
 Ekstr{\"o}m et al. (2012) found a considerable increase
in the main-sequence lifetimes for stellar masses between about 1~\msol and 
2~\msol, by about 25\%, once rotation is included. The new stellar 
evolution modeling, therefore, leaves little doubt that internal rotation 
does indeed prolong main-sequence lifetimes. It seems safe to assume 
that further progress in development of models featuring 
rotational mixing will fully explain large  differences between the lifetimes
of normal and overluminous stars discussed in this paper. 

Stars' differential rotation is quite likely responsible
 not only for longer lifetimes but 
also for stronger X-ray emission of overluminous stars. 
One may speculate that angular momentum transfer from 
the rapidly rotating core and inner layers of a star upwards would 
counteract the secular decay of rotation in the upper layers, 
which would ensure 
that the rotationally induced magnetic field remains strong and the associated
chromospheric activity, hence X-ray luminosity, remains high. 
The stronger X-ray emission 
of old overluminous stars implies, in fact, that any such mechanism 
must be capable of not just maintaining but enhancing that emission
as the star ages.

\section{Conclusions}

The observational data that became available during the last two decades
 reveal that the population of solar-type stars in the mass range 
$M \sim 1$~to~$1.5 \msol$, 
represented mostly by F stars, is non-uniform 
 in terms of stars' stellar evolution rate 
and also in terms of the direction of the evolution of coronal activity. 
First, F stars that have been found to be too luminous for their temperature
and surface gravity -- the {\it overluminous} stars -- are much 
older than estimated from the
standard stellar evolution models. Second, the X-ray emission of old 
{\it overluminous} stars
is on average stronger than that of young stars, which is 
opposite to the expectations from the commonly accepted evolutionary path 
of coronal activity for stars in the solar-mass range. 

These two findings  suggest that standard models of 
stellar evolution and evolution
of stellar activity overlook some crucial physical processes.
Recent developments in stellar evolution modeling that include inner 
rotation and angular momentum transfer as a new factor additional to
mass and chemical composition seem to provide the necessary parameters
and mechanisms to explain the existence of overluminous stars and their
properties. This is good news. What is not so good is that 
computing stellar age from comparison of model calculations and 
observational data is now becoming a task much more challenging 
than before.


\begin{references}

\reference{yale99}
Chaboyer, B., Green, E. M., \& Liebert, J. 1999, AJ, 117, 1360

\reference{eggenbergeriau2010}
Eggenberger, P. Maeder, A. \& Meynet, G. 2010, Proc. IAU Symp., 268, 381  

\reference{eggenbergeraap2010}
Eggenberger, P. et al. 2010, \aap, 519, 116 

\reference{ekstrom}
Ekstr{\"o}m, S., et al. 2012, \aap, 537, 146

\reference{esa97}
ESA, 1997, The Hipparcos and Tycho catalogues,
ed. M. A. C. Perryman (ESA SP-1200; Noordwijk: ESA)

\reference{griffin03}
Griffin, R. F.  \& Suchkov, A. A. 2003, \apjs, 147, 103

\reference{hauck85}
Hauck, B. \&\ Mermilliod, M. 1998, A\&AS, 129, 43

\reference{holmberg09}
Holmberg, J., Nordstrom, B., \& Andersen, J. 2009, A\&A, 501, 941

\reference{micela88}
Micela, G. Sciortino, S., Vaiana, G. S., et al.  1988,  \apj, 235, 798

\reference{pamyat}
Pamyatnykh, A.A., Handler, G., \& Dziembowski, W. A. 2004, \mnras, 350, 1022
 
\reference{stauffer94}
Stauffer, J. R., Caillault, J.-P., Gagn\'e, M., et al. 1994, \apjs, 91, 625

\reference{suchkov00}
Suchkov, A. A. 2000, \apjl, 535, L107 

\reference{suchkov01a}
Suchkov, A. A. 2001, \aap, 365, 554

\reference{suchkov99}
Suchkov, A. A. \&\ McMaster, M. 1999, ApJ, 524, L99

\reference{suchkov03}
Suchkov, A. A., Makarov, V. V., \& Voges, W. 2003, ApJ, 595, 1206

\reference{voges99}
Voges, W., et al. 1999, \aap, 349, 389

\reference{voges00}
Voges, W., et al. 2000, IAU Circ. 7432
\end{references}
\end{document}